\documentclass[%
 reprint,
 amsmath,amssymb,
 aps,
floatfix,
]{revtex4-2}

\usepackage{graphicx}
\usepackage{dcolumn}
\usepackage{bm}
\usepackage{hyperref}

\newif\ifnofigure
\nofigurefalse 
\ifnofigure \renewcommand{\includegraphics}[2][]{}
\fi
\usepackage{lipsum}
\usepackage{booktabs}
\usepackage{physics} 
\usepackage{amsmath, amssymb}

\begin{document}

\title{Deep-learned orthogonal basis patterns for fast, noise-robust single-pixel imaging}

\author{Ritz Ann Aguilar}
\email{raguilar@nip.upd.edu.ph}
\affiliation{%
National Institute of Physics, University of the Philippines Diliman, Quezon City 1101, Philippines\\
}%

\author{Damian Dailisan}
\email{ddailisan@aim.edu}
\affiliation{
 Analytics, Computing, and Complex Systems Laboratory, Asian Institute of Management, Makati City 1229, Philippines\\
}%

\date{\today}

\begin{abstract}
    Single-pixel imaging (SPI) is a novel, unconventional method that goes beyond the notion of traditional cameras but can be computationally expensive and slow for real-time applications. 
Deep learning has been proposed as an alternative approach for solving the SPI reconstruction problem, but a detailed analysis of its performance and generated basis patterns when used for SPI is limited.
We present a modified deep convolutional autoencoder network (DCAN) for SPI on 64$\times$64 pixel images with up to 6.25\% compression ratio and apply binary and orthogonality regularizers during training.
Training a DCAN with these regularizers allows it to learn multiple measurement bases that have combinations of binary or non-binary, and orthogonal or non-orthogonal patterns.
We compare the reconstruction quality, orthogonality of the patterns, and robustness to noise of the resulting DCAN models to traditional SPI reconstruction algorithms (such as Total Variation minimization and Fourier Transform).
Our DCAN models can be trained to be robust to noise while still having fast enough reconstruction times ($\sim$3\,ms per frame) to be viable for real-time imaging.
\end{abstract}

\maketitle

\section{\label{sec:level1}Introduction}


Single-pixel imaging (SPI) uses a single detector to image a scene illuminated with a series of structured light patterns generated by a spatial light modulator (SLM) \cite{Duarte2008}.
The detector measures the corresponding overall intensities from the patterns that are part of a random basis, or an orthogonal basis set such as 2-D Discrete Hadamard Transform (DHT) \cite{Hahamovich2021, sun2017russian} or Discrete Fourier Transform (DFT) \cite{Zhang2015, Zhang2017, Tang2022}.
Depending on the acquisition approach, the image of the scene can be efficiently reconstructed using orthogonal transforms or compressive sensing (CS) techniques~\cite{donoho2006,candes2006compressive,candes2006near}.

One disadvantage of CS methods is their high computational complexity because they employ non-linear iterative schemes to solve an undetermined set of equations.
Alternatively, orthogonal transform methods offer faster reconstruction time with the availability of fast image transform algorithms e.g. Fast Fourier Transform (FFT), but require the scanning of all coefficients of a basis set to perfectly reconstruct the scene.
However, an adaptive sampling path may be established when using orthogonal measurement bases so as to reduce the compression ratio when used for SPI \cite{Xu2021}.
Slow acquisition and reconstruction times of SPI methods present challenges to achieve real-time imaging.  
Overcoming these will make SPI a viable technique for applications such as non-visible imaging \cite{edgar2015simultaneous, gibson2017real, peng2018micro} and, in some cases, even offer a cost-competitive alternative to conventional imaging \cite{Salvador-Balaguer2018, Aguilar2019, Aguilar2022}.

A workaround through deep learning (DL) was proposed by Higham et al. \cite{Higham2018} by using a deep convolutional autoencoder network (DCAN).
They designed the encoder part of the DCAN to simulate the acquisition process of SPI, while the decoding architecture uses convolutional layers to solve the inverse SPI problem.
The DCAN learns optimal binary patterns that can be used as a measurement basis.
While training a DCAN involves a lot of data and computation time, a trained decoder produces high-quality reconstructions at a fast recovery rate, making it capable of real-time imaging.
They pointed out, however, that the learned basis is non-orthogonal; thus, we take particular interest in further exploring this idea.
These learned weights were investigated by modifying the regularization function and consequently doing an orthogonality test.

\begin{figure*}[t]
\centering
\includegraphics[width=\textwidth]{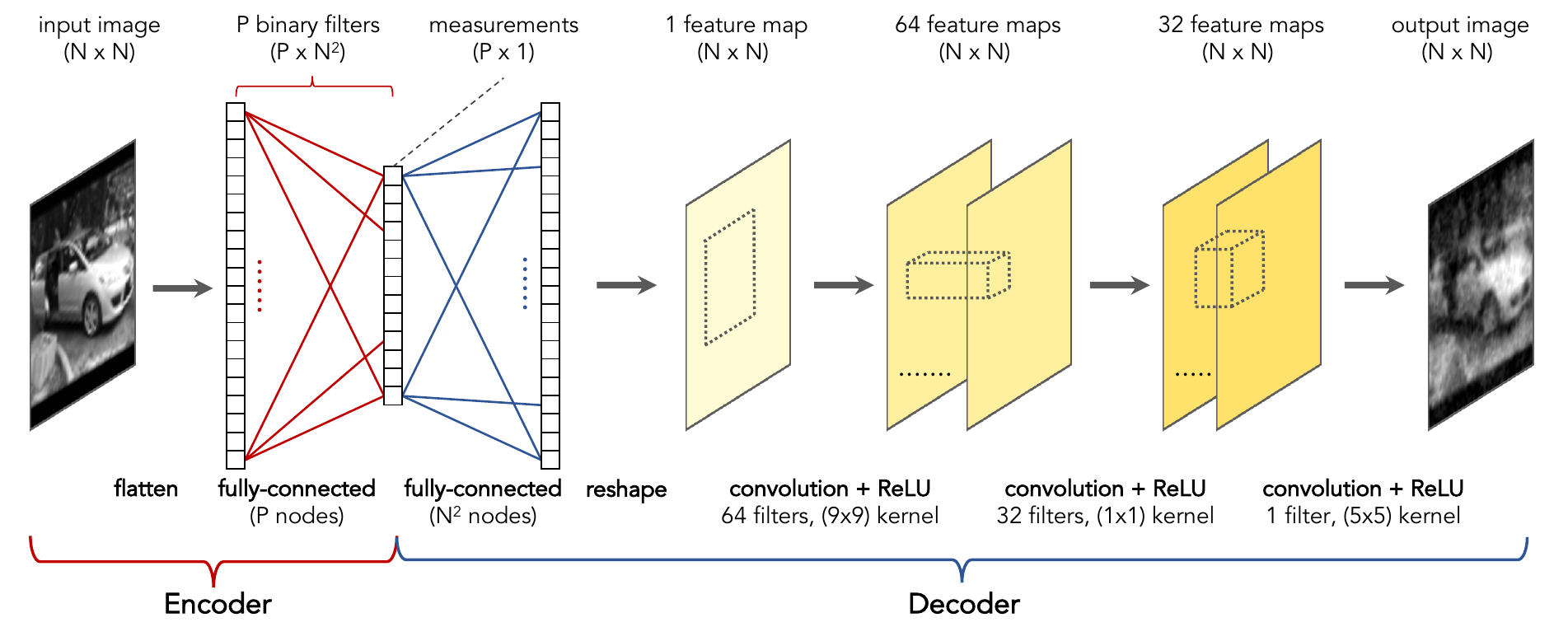}
\caption{\label{fig:dcan} Architecture of the deep convolutional autoencoder.
The encoder architecture mimics single-pixel imaging; modulation patterns are represented by the weights of the fully connected network.}
\end{figure*}

In this paper, we implement an orthogonal regularization function on top of the binary weights regularizer and test if the DCAN can learn an orthogonal basis.
We extend this further by asking the question: \textit{can the network learn an optimized non-binary (orthogonal or not) basis?}
Instead of regularizing the weights, this is achieved by directly imposing constraints on the cost function to learn weights within a certain range of values.
If so, we can have optimal bases, binary or non-binary, that represent the measurement bases we typically use in SPI.
In actual experiments, the learned weights from the encoder can be quantized (or discretized) before feeding them to the decoder.
Note that binary basis here is defined as having coefficients with only two intensity values (e.g. DHT) while non-binary basis has coefficients with more than two (e.g. DFT).

We trained four DCANs to learn four kinds of measurement patterns: non-orthogonal binary or non-binary and orthogonal binary or non-binary.
It is also important that SPI acquisition and recovery methods are robust to various degrees of measurement noise;
hence, we also compare reconstructions of DCANs with state-of-the-art SPI-based CS and orthogonal transform techniques in the presence and absence of noise in measurements.

\section{Single-pixel Imaging Model}
\label{sec:SPI}

The SPI scheme can be modeled as a linear process described by
\begin{equation}
    \vb{y=\Phi x},
    \label{eq:spi}
\end{equation}
where $\vb{\Phi} \in \mathbb{R}^{P\times N^2}$ is the pattern modulation matrix with $P$ modulation patterns of size $N\times N$ pixels, $\vb{x} \in \mathbb{R}^{N^2\times 1}$
represents the (flattened) object scene, and $\vb{y} \in \mathbb{R}^{P\times 1}$ is a 1-D matrix containing the single-pixel measurements~\cite{edgar2019principles,gibson2020single}.
Various SPI methods are used to retrieve $\vb{x}$ using known patterns $\vb{\Phi}$ and measurements $\vb{y}$; however, the cost of taking $P$ measurements may be expensive. 
Researchers have then employed compressive sensing to retrieve $\vb{x}$ from Eq.~(\ref{eq:spi}) using $P \leq N^2$ measurements.
Otherwise, $\vb{x}$ may be treated as a representation of the image in some known basis.

Our work uses two representative reconstruction algorithms as baselines: reconstructions using compressive sensing and image transforms.

\subsection{Compressive Sensing: Total Variation Minimization}
Compressive sensing techniques~\cite{donoho2006,candes2006compressive,candes2006near} use a set of undersampled measurements $\vb{y}$ to solve the linear inverse problem of Eq.~(\ref{eq:spi}). 
This results in an underdetermined linear system, which has a solution, provided sparsity priors on the target signal.
For natural images, sparsity may arise from the gradient of the image~\cite{krahmer2017total}.
The optimization problem is 
\begin{equation}
\begin{aligned}
    \min \quad &\|\vb{z}\|_{l_1}, \\
    s.t. \quad &\vb{z = \Psi x}, \\
     &\vb{y = \Phi x},
\end{aligned}
\label{eq:tv_min}
\end{equation}
where $\vb{\Psi}$ is a basis transform matrix and $l_1$ norm is used to obtain the total variation (TV) of the image.

For TV minimization, one can write $\vb{\Psi=\Theta}$ where $\vb{\Theta}$ is a gradient calculation matrix.
The above minimization can then be solved using an augmented Lagrange multiplier (ALM) method~\cite{li2013efficient}, as implemented in~\cite{Bian2018}.

\subsection{Image Transforms: Discrete Fourier Transform }
Fourier transforms are commonly used in signal processing applications to represent signals as linear combinations of sines and cosines.
For SPI, the DFT takes advantage of sparsity in the Fourier spectrum of images and that most of their information is concentrated in low spatial frequencies~\cite{Taubman2012, Bian2016}.

Fourier single-pixel imaging~\cite{Zhang2015} uses sets of sinusoidal patterns with $K$ patterns for each step and an $K$-step algorithm to reconstruct the Fourier spectrum from SPI measurements.
Patterns in each set are offset by integer multiples of a phase shift $\Delta\psi$.
The intensity of the $k$\textsuperscript{th} patterns is  
\begin{equation}
    \Phi_k\left(m, n;\mu,\nu\right) = a + b\cos{\left[ 2\pi \left(\mu m + \nu n \right) + k\Delta\psi \right]},
\end{equation}
where $(m, n)$ are the pixel coordinates with corresponding spatial frequency coordinates $(\mu, \nu)$, $a$ is the DC term, $b$ is the sinusoidal amplitude. 

For $K=2$, we can use the two-step algorithm with a phase shift of $\Delta \psi=\pi/2$ ($k=0,1$). 
We then obtain the following set of intensity equations,
\begin{align}
\begin{cases}
    &\Phi_0 = a + b\cos\theta(m, n), \\
    &\Phi_1 = a + b\sin\theta(m, n),
\end{cases}
\end{align}
with $\theta(m, n) = 2\pi(\mu m + \nu n)$.
A projected pattern interacts with the object scene, which reflects light with an intensity
\begin{equation}
   G_k(\mu,\nu) = \iint X(m, n) \Phi_k \left(m, n;\mu,\nu \right) dm\,dn,
   \label{eq:field}
\end{equation}
where $X(m, n)$ represents reflectivity of the object.
This light collected by a single-pixel detector measurement can then be expressed as $y_k(\mu,\nu) \propto G_k(\mu,\nu)$.
These measurement terms can be combined to retrieve the Fourier spectral coefficients,
\begin{align}
\begin{split}
H(\mu,\nu) &= y_0 - jy_1,\\
&= b \iint X(m, n) \exp\left(-j\theta(m, n) \right)dm\,dn ,\\
&\propto \mathcal{F}\{X(m, n)\}.
\end{split}
\end{align}
Thus, the measurements of the various sets of measurement patterns can be interpreted as a Fourier transform of the object, $X(m, n)$, and the object scene may be recovered by taking the inverse Fourier transform of the coefficients $G(\mu,\nu)$ derived from the single-pixel measurements.

\section{Deep learning Architecture for Single-pixel Imaging}
\label{sec:DL_SPI}

\subsection{Deep Convolutional Autoencoder}
We used a deep convolutional auto-encoder network (DCAN) architecture described in \cite{Higham2018}, where the encoding architecture mimics the pattern modulation process in single-pixel imaging (see Fig.~\ref{fig:dcan}).
The DCAN is trained to replicate an $N\times N$ input scene as an output scene of the same dimensions.
Once trained, we can recover filter patterns from the encoder and use the decoder to reconstruct an image from single-pixel measurements.

The encoder uses a single fully connected layer with $P$ neurons and a linear activation function, which results in the operation
\begin{equation}
    y_{j,\mathrm{enc}}(\vb{x}) = \vb{w}_j^\intercal
    \vb{x} + b_j
    \label{eq:dcan_dense}
\end{equation}
for each neuron $j \in \{1,2,\dots,P\}$ in the network.
Here, the $P$ weight vectors $\vb{w}_j$ correspond to $P$ flattened filter images.
With $b_j=0$, the encoder outputs a vector of $P$ measurements corresponding to the total intensity of the input scene modulated by the filters.

While our computational model learns weights $w \in [-1,1]$ --- deep learning algorithms typically learn better when weights are zero-centered --- experimental devices that modulate light are better represented by $w_\mathrm{exp} \in [0,1]$.
This disparity in representing $w$ and $w_\mathrm{exp}$ is resolved by taking the difference between the single-pixel measurements that use the filter pattern $\vb{w}_{j,\mathrm{exp}}$ and its one's complement:
\begin{align}
    y_{j,\mathrm{exp}}(\vb{x}) =& \vb{w}_{j,\mathrm{exp}}^\intercal
    \vb{x} - \left(1-\vb{w}_{j,\mathrm{exp}}^\intercal\right)
    \vb{x}\\
    =& \frac{1}{2}\left(1+\vb{w}_j^\intercal\right)
    \vb{x} - \frac{1}{2}\left(1-\vb{w}_j^\intercal\right)
    \vb{x}\\
    =& \vb{w}_j^\intercal \vb{x}
\end{align}
which is analogous to the computational implementation described by Eq.~(\ref{eq:dcan_dense}).

The decoder takes in the single-pixel measurements as input and outputs a reconstruction.
Its first layer upscales the input $P$-dimensional measurement vector to an $N\times N$ image using a fully connected layer with $N^2$ nodes.
More common architectures of convolutional decoders gradually upscale the input after several convolutional and upsampling layers~\cite{rizvi2019improving,sun2020learning,song2020unified,liu2022fitbeat}, but this limits the input size to the decoder, which is the number of measurements $P$.
However, since we vary the number of measurements $P$, we chose to upscale using a fully-connected layer before the convolutional layers to ensure the output will be an $N\times N$ image.
A batch normalization layer~\cite{ioffe2015batch} inserted after the fully connected layer of the decoder speeds up the convergence of training.
The final three layers of the decoder are convolutional layers with kernel dimensions (9,9,1), (1,1,64), and (5,5,32).
Each convolutional layer uses a rectified linear unit (ReLU) activation function.
The DCAN was trained using the Adaptive Moment Estimation (Adam) optimizer for 50 epochs with a mean squared error cost function.
We implemented these using \texttt{Python v3.9} and \texttt{Tensorflow v2.7} running on an NVIDIA GTX 1080Ti.


\subsection{Binary Regularizer}
Higham et al.~\cite{Higham2018} presented a deep learning model that learns an optimal binary basis set with values of $\pm1$.
They used a regularization scheme,
\begin{equation}
    \Omega_\mathrm{binary}= \frac{1}{PN^2}\sum_{j}^{P} \sum_{i}^{N^2}\left(1+w_{i j}\right)^{2}\left(1-w_{i j}\right)^{2},
    \label{eq:binary_regularizer}
\end{equation}
to drive weights towards positive and negative unitary values.
Their work discusses that the above regularization guides the DCAN to an optimized, albeit non-orthonormal, binary basis set.
As aforementioned, this is an interesting case that needs to be further investigated; we perform an orthogonality test on the learned weights, which we discuss in the next section.

\subsection{Orthogonal Regularizer}
The weights for all $P$ neurons of the fully connected network of the encoder can be written as a matrix $\vb{W}=[\vb{w}_1,\vb{w}_2,\dots,\vb{w}_P]$.
However, weight updates in backpropagation provide no guarantee that the individual patterns $\vb{w}_j$ have similar magnitudes.
Hence, we can define a normalized weights matrix $\vb{\tilde{W}}$ where each pattern $j$ is normalized as $\vb{\tilde{w}}_j=\vb{w}_j/\|\vb{w}_j\|$.
With this, we can impose an orthogonal regularization function on top of the binary weights regularizer in Eq.~(\ref{eq:binary_regularizer}) of the form
\begin{equation}
    \Omega_\text{ortho} = \frac{1}{P^2}\sum\vb{\tilde{W}}^\intercal \vb{\tilde{W}} - \vb{I},
    \label{eq:orthogonal_regularizer}
\end{equation}
where $\vb{I}$ is the identity matrix.
A basis set $\vb{W}$ is orthogonal if $\Omega_\text{ortho}=0$.

\subsection{Training and Testing}
We used the STL--10 dataset~\cite{coates-stl2011} image recognition dataset, which contains 100,000 colored, unlabeled images of animals and vehicles used to develop unsupervised feature learning, deep learning, and self-taught learning algorithms.
All images in the STL--10 dataset are color images of size $96\times96$, but we resized them to $64\times64$ and converted them to grayscale. 
We employed a 9:1 train--test ratio to train and evaluate the models.

We also trained the DCAN in two phases.
First, we trained the entire network without any noise, regularization, and constraints in the encoder.
This prevented the weights of the decoding layers from getting trapped in a local optima early in the training process~\cite{Higham2018} and gave us a common decoder to use with different encoding schemes.
Once this initial training was finished, the decoding layer weights were fixed and the encoding layers were re-initialized and retrained with noise, regularization, and constraints.
Additionally, Gaussian noise was added to the encoder measurements to further regularize the deep learning network.

\begin{figure*}[t]
\centering
\includegraphics[width=\linewidth]{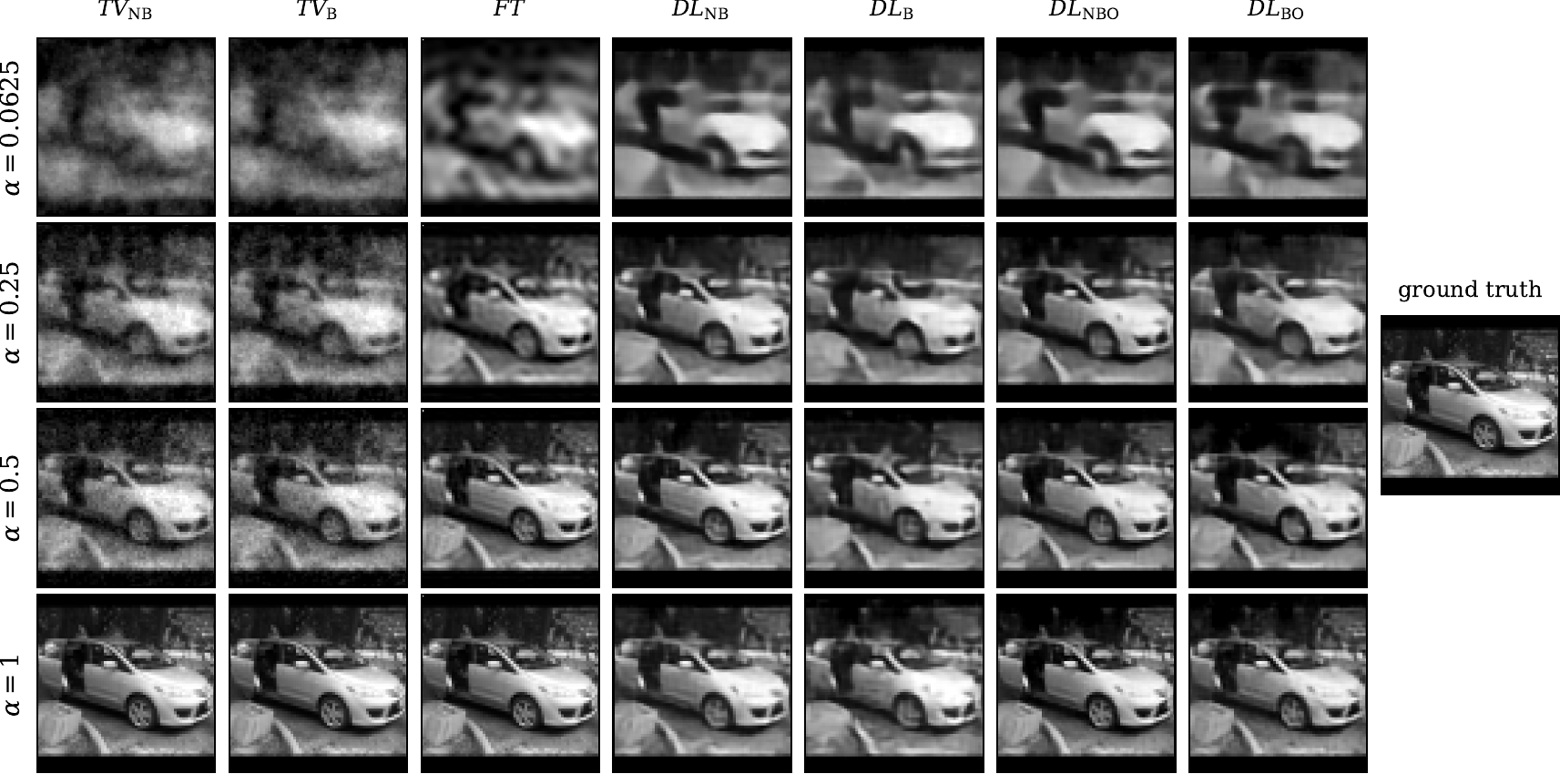}
\caption{\label{fig:images_sample} Sample image reconstructions of the different models (columns) at different sampling ratios $\alpha$ (rows).
The FT result at $\alpha = 1$ serves as the ground truth following a perfect reconstruction.
At a low sampling ratio $\alpha=0.0625$, TV results are barely recognizable.
In general, using TV results in noisy reconstructions for $\alpha<1$.}
\end{figure*}

\section{Results and Discussion}
\label{sec:results}

We simulate image reconstructions for three different methods: Total Variation (TV) minimization, Fourier Transform (FT), and Deep Learning (DL). 
For DL, we split it further into two categories: each trained with and without an orthogonal regularizer with constraining and regularizing functions described in Sec. \ref{sec:DL_SPI}.
Moreover, we explored using both binary and non-binary patterns for TV and DL to include patterns that are typically used in most SPI experiments. 
This resulted in seven algorithms, namely: (i) $TV_\mathrm{B}$, (ii) $TV_\mathrm{NB}$, (iii) FT, (iv) $DL_\mathrm{B}$, (v) $DL_\mathrm{BO}$, (vi) $DL_\mathrm{NB}$, and (vii) $DL_\mathrm{NBO}$, where the subscripts B, BO, NB, and NBO stand for binary, binary orthogonal, non-binary, and non-binary orthogonal, respectively.

We then tested and compared the algorithms for various sampling ratios and noise levels to check their robustness and efficiency. 
As a benchmark for possible applications to real-time imaging, we also analyze the interplay between computational complexity and sampling ratios of the different models. 
Our focus is on the reconstruction rates, as while training of the DL approaches typically takes a long time, it can be done once before the use of the models.

\subsection{Sampling Ratio} 

Image reconstructions of the seven algorithm for different sampling ratios $\alpha$  are shown in Fig.~\ref{fig:images_sample}.
Sampling ratio is defined as $\alpha=P/N^2$, or the ratio of measurements to the total number of pixels in an image.
We set the image resolution to $N=64$ pixels in our simulations.

Reconstructions using TV have the lowest quality and appear visibly noisy for $\alpha<1$.
On the other hand, the FT and DL methods have reconstructed the image well even at much lower sampling ratios ($\alpha\leq 0.25$).
However, the FT approach has ringing artifacts, while the DL approach takes on a smoother appearance.

\begin{figure}[tb]
\centering
\includegraphics[width=\linewidth]{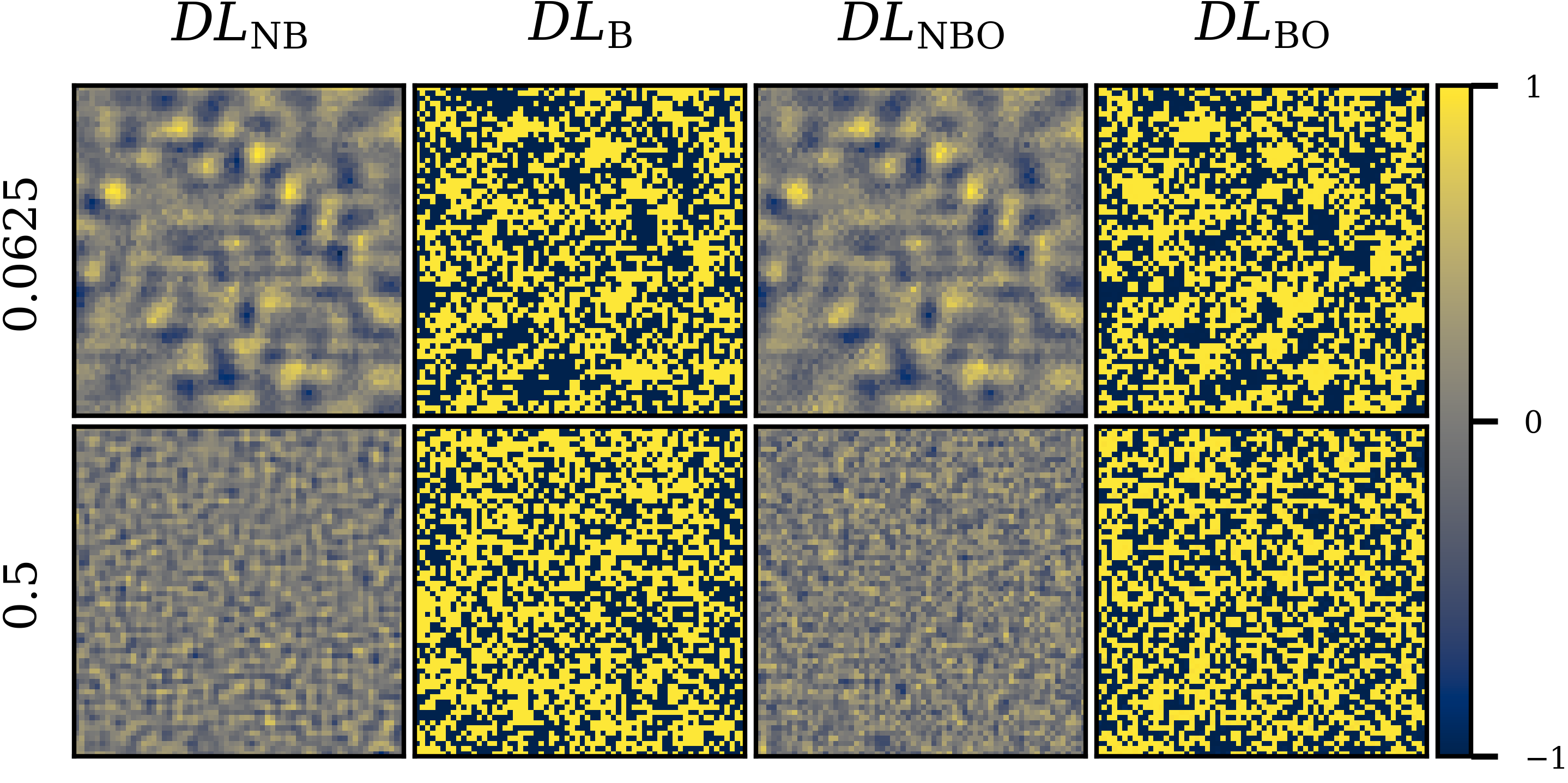}
\caption{\label{fig:learned_weights} Learned basis patterns of the DCAN. Higher sampling ratios $\alpha$ correspond to more patterns $P$, which in turn influences the spatial frequency of learned patterns. With less patterns $P$, the DCAN learns patterns with larger blobs, corresponding to a lower spatial frequency pattern.}
\end{figure}

We attribute the smoothed appearance of reconstructions at low sampling rates to low spatial resolution patterns for the FT and DL methods.
Inspection of the learned weights of the DL approach shows learning prioritized low frequency patterns when only a few patterns are used (Fig.~\ref{fig:learned_weights}). Similarly, the FT method provides recognizable reconstructions at low sampling ratios by exploiting the fact that most information for natural images are concentrated at low spatial frequencies.
In particular, we set its sampling path to an outward spiral manner that samples the low frequencies in the Fourier domain producing spectra similar to Ref.~\cite{Aguilar2019}.

\subsection{Orthogonality}

With constraints imposed and regularizers added to the DL model, the encoder learns different modulation patterns for measurements.
The sample weights learned by the DL models for two different sampling ratios, 0.0625 and 0.5, are shown in Fig.~\ref{fig:learned_weights}.
As expected, the $DL_\mathrm{B}$ and $DL_\mathrm{BO}$ have learned weights with only two intensity values and the patterns have more pixel density --- higher spatial resolution --- at $\alpha = 0.5$ compared to $\alpha = 0.0625$.
However, texture-wise, the resulting $DL_\mathrm{B}$ patterns, in particular, differ from that of Higham et al.'s.
Their patterns resemble a blob-like structure and appear smoother than our $DL_\mathrm{B}$ patterns.

With the same initialization, the $DL_\mathrm{NB}$ and $DL_\mathrm{NBO}$ models have similar spatial variation patterns but differ in terms of range of intensity values.
While weights were constrained to values of range $[-1,1]$, $DL_\mathrm{NB}$ can utilize a wider range of values compared to $DL_\mathrm{NBO}$, which manifests as a washed out and noisier appearance.
The addition of an orthogonal regularizer for non-binary patterns prevents the DCAN's weights from utilizing the full range of $[-1,1]$; this may present problems when using these patterns on an SLM.

\begin{figure}[tb]
\centering
\includegraphics{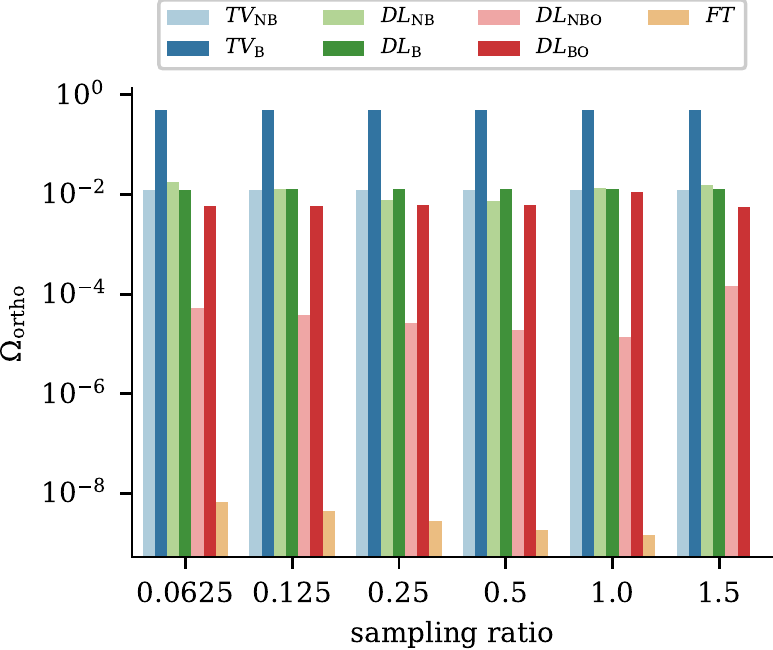}
\caption{\label{fig:orthogonality} Orthogonal regularizer score for the different SPI models. The FT basis pattern should have a theoretical $\Omega_\mathrm{ortho}=0$ as it uses an orthonormal basis set, but we obtain a small, non-zero value due to quantization. The $DL_\mathrm{NBO}$ approach resulted in the lowest $\Omega_\mathrm{ortho}$, but the other approaches already have small $\Omega_\mathrm{ortho}$, with the exception of $TV_\mathrm{B}$.}
\end{figure}

For a more qualitative analysis of the set of patterns learned by the DCAN, we plotted the orthogonality score $\Omega_\mathrm{ortho}$ (Eq.~\ref{eq:orthogonal_regularizer}) as a function of the sampling ratio for the seven algorithms (see Fig.~\ref{fig:orthogonality}).
Note that we used uniform random binary and non-binary patterns as measurement bases for the TV minimization algorithm.
Logarithmic scaling of the $y$-axis emphasizes the differences in $\Omega_\mathrm{ortho}$.
We set the FT method as a baseline algorithm with an orthogonality score almost equal to zero ($\Omega_\mathrm{ortho}<10^{-8}$) for all sampling ratios.
The sampling ratio $\alpha = 1.0$ completely captures all Fourier coefficients, hence FT cannot have oversampled measurements.
Theoretically, it should have a $\Omega_\mathrm{ortho} = 0$ but with the patterns quantized, we get a non-zero value, albeit very small.

Observing the plots in Fig.~\ref{fig:orthogonality}, the algorithm with the highest orthogonality score is $TV_\mathrm{B}$ ($\Omega_\mathrm{ortho}\approx 10^{-1}$) followed by $DL_\mathrm{NB}$ ($\Omega_\mathrm{ortho}<10^{-1}$).
The one with the least orthogonality score, excluding FT, is $DL_\mathrm{NBO}$ with $\Omega_\mathrm{ortho}<10^{-4}$ for all sampling ratios except at $\alpha > 1$, which already has repeating patterns.
This is followed by $DL_\mathrm{BO}$ with $\Omega_\mathrm{ortho}<10^{-2}$, 
which is expected since $DL_\mathrm{NBO}$ and $DL_\mathrm{BO}$  have orthogonal regularizers imposed on them.
However, $DL_\mathrm{BO}$ does not achieve as low a score as $DL_\mathrm{NBO}$.
The binary and orthogonal regularizers compete with each other, with a greater emphasis on binarization.
Once the weights converge to a binary set, it gets trapped in a local minima where no further improvements to orthogonality can be done.
The DL methods without orthogonal regularizer, $DL_\mathrm{NB}$ and  $DL_\mathrm{B}$, still have low orthogonality score at $\Omega_\mathrm{ortho}\approx 10^{-2}$.

While not orthogonal, $TV_\mathrm{NB}$, $DL_\mathrm{B}$, $DL_\mathrm{NB}$, and $DL_\mathrm{NBO}$ have low enough $\Omega_\mathrm{ortho}$.
Using an orthogonal basis set does not guarantee good reconstruction at a low sampling ratio, as in the case of $TV_\mathrm{NB}$ ($\Omega_\mathrm{ortho} \approx 10^{-2}$).
Orthogonality needs to be coupled with a reconstruction algorithm that can take advantage of the efficient orthogonal representations of images, which deep learning can achieve.

\subsection{Noise} 
\label{sec:noise}

\begin{figure*}[!tb]
\centering
\includegraphics[width=0.9\linewidth]{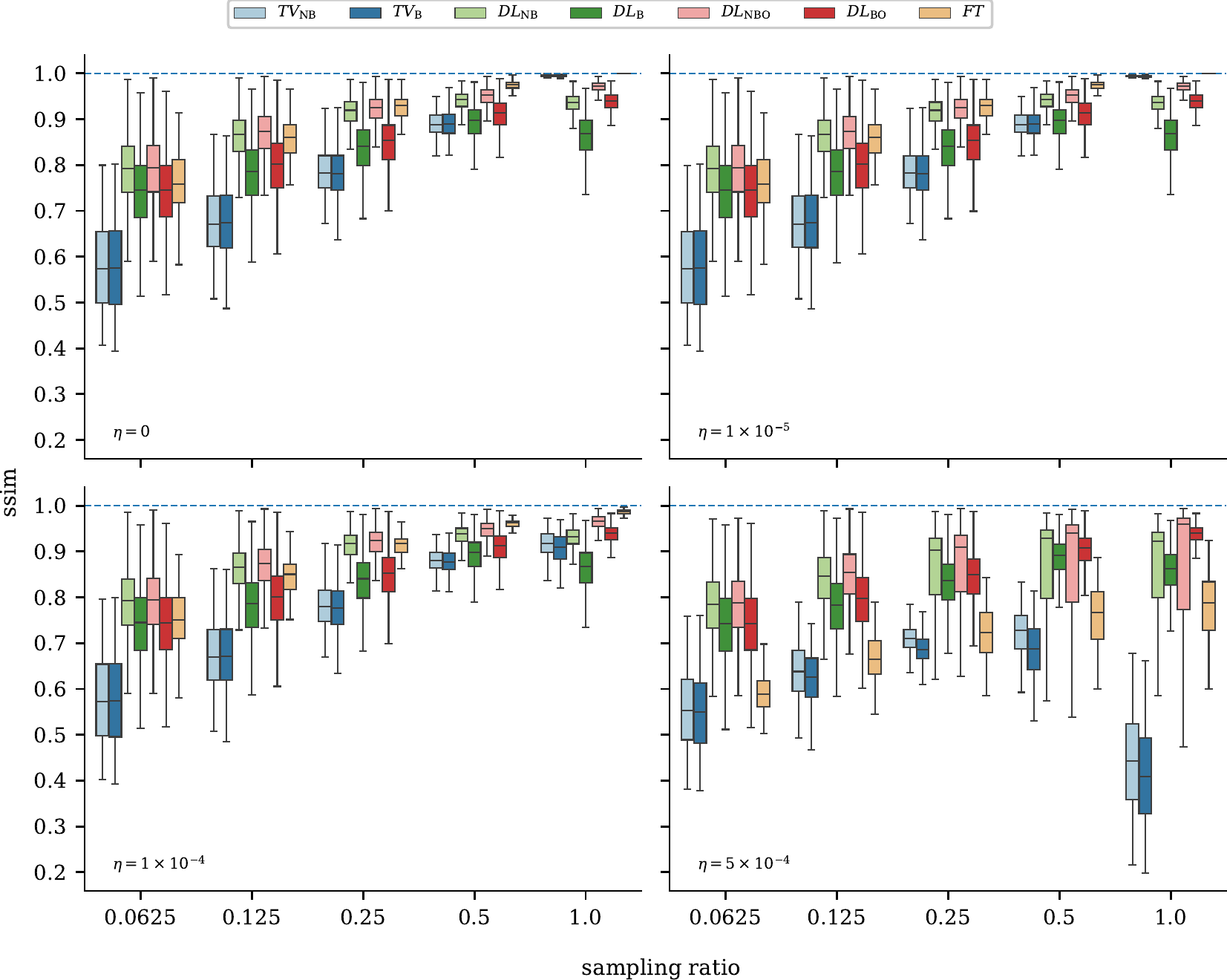}
\caption{\label{fig:noise_ssim} Analysis of measurement noise and reconstruction quality (using SSIM) for different noise levels $\eta$. Reconstruction quality is significantly affected at noise levels near $\eta=5\times10^{-4}$.}
\end{figure*}

In the previous sections, we neglected the effects of noise; however, experimental measurements are 
subjected to various noise sources, such as ambient light and current fluctuations.
This section studies each method's robustness to measurement noise.
We assume Gaussian white noise at the detector level, which follows a probability distribution $P(n)=\frac{1}{\sqrt{2\pi}\sigma}\exp\left(-\frac{n^2}{2\sigma^2}\right)$, where $n$ is the measurement noise and $\sigma$ is the standard deviation.
We vary the noise level $\eta=\sigma/N^2$ to obtain different values of $\sigma$ and run simulations using images sampled from the remaining 10,000 test images.

\begin{figure*}[!t]
\centering
\includegraphics[width=\linewidth]{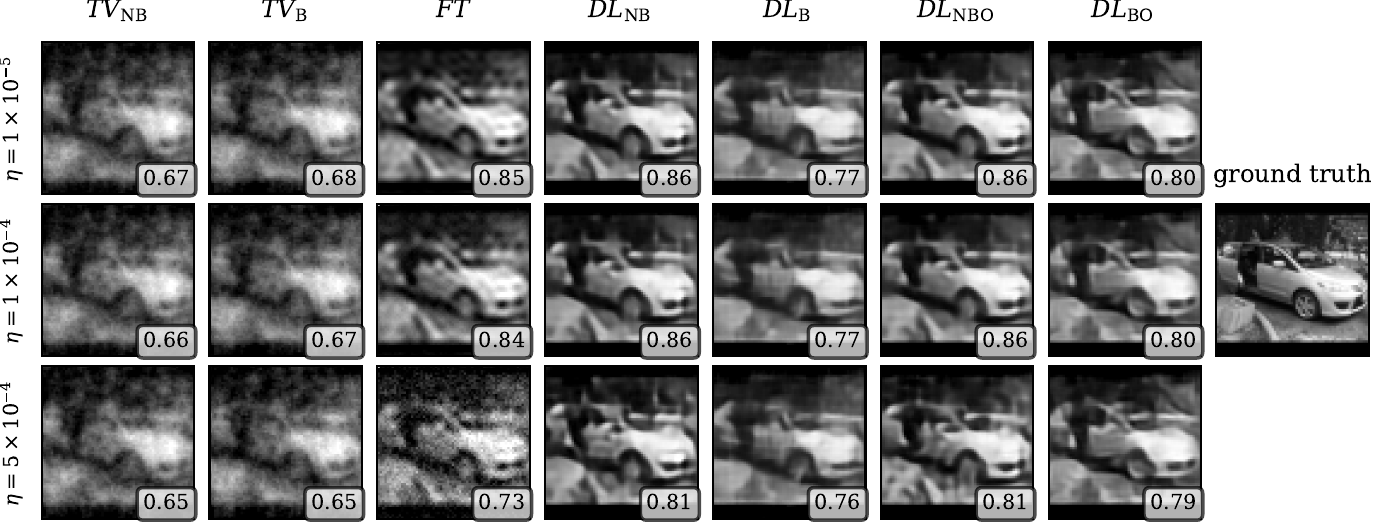}
\caption{\label{fig:noise_sample} Sample image reconstructions of the different models (columns) at different noise levels $\eta$ (rows) with measurement noise.
Images are reconstructed using a sampling ratio of $\alpha=0.0625$.
At the lower right corner of each image is the SSIM score to quantify the reconstruction quality.
}
\end{figure*}

For different sampling ratios, we measured the structural similarity index (SSIM)~\cite{Wang2004} as a quantitative metric of human perception (Fig.~\ref{fig:noise_ssim}).
Without noise, the FT and DL methods have higher image quality than TV methods for sampling ratios $\alpha<1$.
However, when $\alpha=1$, both TV and FT approaches have near-perfect SSIM values.
The non-binary patterns also result in better reconstructions across all sampling ratios for the deep learning methods.
At a low sampling ratio ($\alpha=0.0625$), 
we plot the image reconstruction of each single-pixel imaging model at varying noise levels (see Fig.~\ref{fig:noise_sample}) with the corresponding SSIM value for each tabulated in Table~\ref{tab:noise_ssim}.
We observe higher reconstruction quality for the FT and DL approaches.
Noise greatly affects the quality of images reconstructed by TV for both binary and non-binary patterns; there is not much high spatial frequency details.
On the other hand, FT has a grainier appearance at high noise levels while DL models have patch-like artifacts.

\begin{table*}[tb]
\begin{center}
\begin{minipage}{\textwidth}
\caption{\label{tab:noise_ssim} Corresponding structural similarity index (SSIM) values of the sample image reconstructions of the different models in Fig.~\ref{fig:noise_sample}. The best values are in bold.} \label{tab2}
\begin{tabular*}{\textwidth}{@{\extracolsep{\fill}}lccccccc}
\toprule%
& \multicolumn{7}{@{}c@{}}{Single-pixel Imaging Models} \\\cmidrule{2-8} %
Noise Level & $TV_\mathrm{NB}$ & $TV_\mathrm{B}$ & $FT$ & $DL_\mathrm{NB}$ & $DL_\mathrm{B}$ & $DL_\mathrm{NBO}$ & $DL_\mathrm{BO}$ \\
\midrule
$1\times10^{-5}$  & 0.67 & 0.68 & 0.85 & \textbf{0.86} & 0.77 & 0.86 & 0.80 \\
$1\times10^{-4}$  & 0.67 & 0.67 & 0.84 & \textbf{0.86} & 0.77 & 0.86 & 0.80 \\
$5\times10^{-4}$  & 0.64 & 0.65 & 0.74 & 0.81 & 0.75 & \textbf{0.81} & 0.78 \\
\botrule
\end{tabular*}
\end{minipage}
\end{center}
\end{table*}

The added noise reduces the reconstruction quality of TV and FT methods and is more apparent at high sampling ratios.
At the highest noise level ($\eta=5\times10^{-4}$), the reconstruction quality of both TV and FT methods falls below those of the deep learning methods.
Deep learning methods also have reduced image quality due to noise, most prominently with the $DL_\mathrm{NBO}$, but the rest of the deep learning methods have better reconstruction quality than traditional SPI methods.
The B, BO, and NB methods are most robust to high noise levels, with the orthogonalized patterns reducing the variance of SSIM values compared to the non-orthogonalized basis patterns.
The DL models with non-binary patterns ($DL_\mathrm{NB}$ and $DL_\mathrm{NBO}$) have generally better reconstruction quality even at all noise levels than their binary counterparts ($DL_\mathrm{B}$ and $DL_\mathrm{BO}$).
However, we also observe a large spread of SSIM values for the non-binary DL methods at higher sampling ratios.

For practical considerations, deep learning models that use the DCAN architecture are best suited for undersampled ($\alpha<0.5$) SPI reconstructions.
In such scenarios, deep learning (notably the NB and NBO patterns) methods have similar reconstruction quality with FT for low measurement noise and are more robust to noisy measurements.
Large variances at high sampling ratios may be due to insufficient training of the deep learning methods; large network sizes limited the training epochs at high sampling ratios.
High measurement noise levels also reduce image quality at high sampling ratios as more noisy measurements make it harder to correlate basis patterns with SPI measurements.
To best mitigate the effects of noise, multiple measurements can be taken and averaged for each basis pattern, but this can be infeasible for real-time SPI applications.







\subsection{Reconstruction Time}

\begin{figure}[tb]
\centering
\includegraphics{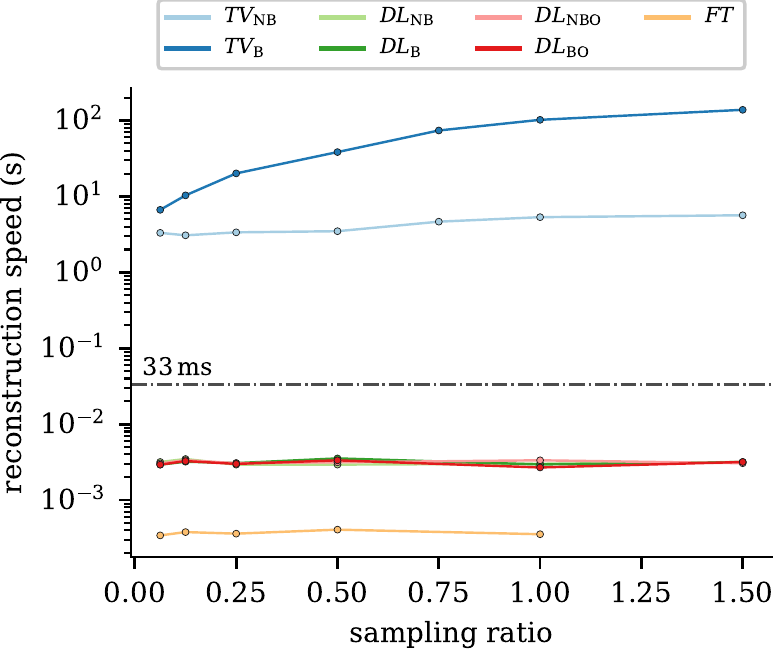}
\caption{\label{fig:reconstruction_rate} Reconstruction times of the different algorithms for different sampling ratios. The gray dashed line marks reconstruction speeds of 33\,ms (or 30\,fps). Both TV methods have the slowest times and increase with sampling ratio, while the DL and FT methods maintain fast reconstruction times even with increasing sampling ratios. }
\end{figure}

Single-pixel imaging methods face challenges on two fronts: acquisition and reconstruction times.
Combined, these two factors determine the frame rate achieved by any SPI method.
Acquisition times 
are primarily dependent on the hardware used to project patterns (SLM) and detect measurements (bucket detectors), while reconstruction times vary with the algorithm used.
We compare reconstruction times for the different algorithms to show which ones can be used for real-time SPI applications (Fig.~\ref{fig:reconstruction_rate}).

Total Variation algorithms perform the slowest because of the non-linear iterative nature of image reconstruction.
As no orthogonality between patterns is exploited, increasing the sampling ratio further increases reconstruction time and reduces the frame rate.
The FT method, on the other hand, performs measurements to reconstruct the frequency domain representation and computes the inverse to retrieve the reconstructed image.
Fourier transforms have fast implementations (in the form of Fast Fourier Transform algorithms), which accounts for fast image reconstruction speeds of this approach.
This method is independent of sampling ratio: more samples correspond to more coefficients in the frequency domain, but the total size of the spectrum remains fixed before inversion.

The DCAN-based approaches all have similar reconstruction speeds that also lack a clear dependence on increasing sampling ratio.
While not as fast as FT, these reconstruction times are still around one order of magnitude faster than 33\,ms (or 30\,fps), which leaves some leeway for the acquisition system.
In addition to the noise robustness discussed in Sec.~\ref{sec:noise}, DCAN is a good alternative to FT for real-time, noise tolerant SPI systems.


\section{Conclusions}
Our work demonstrates an approach to learning an orthogonal basis by introducing an orthogonal regularizer in a DCAN.
When used as the sole regularizer for the network, the orthogonal regularizer produces the with the highest orthogonality.
However, when used in conjunction with another regularizer (such as a binary regularizer), the degree of orthogonality of the resulting patterns decreases.
We also find that even without the orthogonal regularizer, the orthogonality of learned patterns is much lower than that of a purely uniform random basis set.
Because orthogonal bases allow for efficient, sparse representation of images, these near-orthogonal bases benefit deep learning SPI and allow for good reconstructions even at low sampling ratios.
While FT is fast and highly efficient, its reconstruction quality degrades in the presence of high measurement noise.
The DL models can be trained to be robust to noise while still having fast enough reconstruction times to be viable for real-time imaging.
Total Variation minimization suffers under high noise levels and also has the slowest reconstruction time at all sampling ratios.

Unlike FT and TV, DCANs take an investment of time and resources when imaging parameters such as resolution of the number of measurements change.
For DCAN SPI, the number of measurement patterns determines the network structure; hence increasing or decreasing this parameter warrants retraining the entire model.
Significant advances in GPU technology and data availability have made deep learning a popular method for various tasks, but training a neural network still takes considerable time and resources, unlike FT or TV, which can be used immediately.
Finite GPU memory also limits the image resolution and measurements.
Further studies into more efficient deep learning architectures applied to SPI can potentially overcome these limitations and increase the versatility of deep learning for SPI.

\section*{Acknowledgments}
D. Dailisan acknowledges computing resource support provided by the Analytics, Computing, and Complex Systems Laboratory at the Asian Institute of Management.

\tiny
\bibliography{revtex}

\end{document}